\documentclass[reprint,amsmath,amssymb,aps,prl]{revtex4-2}
\usepackage{graphicx}
\usepackage{amsmath}
\usepackage{amssymb}
\usepackage{hyperref}
\usepackage{babel}
\usepackage{xcolor}

\newcommand{\ud}{\ensuremath{\mathrm{d}}}

\newcommand{\msun}{\ensuremath{\,\mathrm{M}_\odot}}

\begin{document}
\title[Minimum neutron star mass]{The minimum neutron star mass in neutrino-driven supernova explosions}

\author{Bernhard M\"uller}
\author{Alexander Heger}
\affiliation{School of Physics and Astronomy, Monash University, Clayton, VIC~3800 Australia}
\author{Jade Powell}
\affiliation{Centre for Astrophysics and Supercomputing, Swinburne University of Technology, Hawthorn, VIC 3122, Australia}

\begin{abstract}
Supernova theory has struggled to explain the lightest known neutron star candidate with an accurate mass determination, the $1.174\msun$ companion in the eccentric compact binary system J0453+1559. To improve the theoretical lower limit for neutron star birth masses, we perform 3D supernova simulations for five stellar models close to the minimum mass for iron core collapse. We obtain a record-low  neutron star mass of $1.192\msun$ and a substantial kick of $\mathord{\sim}100\,\mathrm{km}\,\mathrm{s}^{-1}$. Given residual uncertainties in stellar evolution, a neutron star origin for the $1.174\msun$ object remains plausible.
\end{abstract}
\maketitle

\textit{Introduction.}---Compact object masses are among the most critical astrophysical observables because of their implications for high-density nuclear physics, stellar evolution, and supernova explosion physics. In recent years, new records for the maximum neutron star mass \citep{antoniadis_16,romani_22} together with complementary constraints on neutron star radii and tidal deformability from X-ray \citep{miller_21} and gravitational wave \citep{abbott_18_long,de_18b,abbott_20_long} observations, have provided important insights on the nuclear equation of state. 

The \emph{lowest} neutron star mass also has significant implications.  In contrast to the upper mass limit, the lower limit is due the astrophysical formation \emph{path}, and is not a limit inherent to neutron stars and their equation of state.  The best known way to make the lowest possible neutron star masses is the gravitational collapse of the iron 
core of massive stars  
\citep{woosley_15} 
or of the O-Ne-Mg core of Super Asymptotic Giant Branch (SAGB) stars
\citep{nomoto_87,jones_13,leung_19}.
The masses of young neutron stars will reflect the core sizes and compositions in massive stars, although the exact 
neutron star mass will depend on details of the explosion dynamics.
Hence, the lowest neutron star mass is a key parameter for testing our understanding of stellar evolution and nuclear physics \citep{H01} through advanced burning stages to core collapse \citep{WHW02}. 
In the alternative scenario of accretion-induced collapse
\citep{nomoto_91,yoon_04,dessart_06_b},
a higher electron fraction and rotational stabilization likely result in a higher mass of collapsing white dwarf and the resulting neutron star.

Precise measurements of neutron star masses in binary systems have pushed minimum gravitational neutron star mass below $1.2\msun$. The double neutron star system J0453+1559 was found to contain a companion with only $1.174\msun$ \citep{martinez_15}. There is also a candidate neutron star of $1.2\msun$ in J1807-2500B \citep{oezel_16,ridolfi_19}, though the alternative interpretation  of this object being a white dwarf cannot be excluded in this case. A white dwarf origin for the $1.174\msun$ object in J0453+1559 has also been proposed \citep{tauris_19}. There are further claims of even lower neutron star masses \citep{doroshenko_22}, but these are beset with large error bars.

Such low neutron star masses present a challenge for current stellar evolution and supernova explosion models, which, to date, have failed to obtain such low masses. The record has long been held by simulations of electron-capture supernovae with a minimum baryonic neutron star mass of  $1.36\msun$ \citep{huedepohl_10}, which translates into a gravitational mass of about $1.24\msun$, with only small uncertainties from the nuclear equation of state. In this narrow stellar evolution channel for stars around $\mathord{\sim}9\msun$ in zero-age main sequence (ZAMS) mass, dynamical collapse is believed to occur after the formation of an O-Ne-Mg core in SAGB stars due to rapid electron capture (EC) on ${}^{20}\mathrm{Ne}$ and ${}^{24}\mathrm{Mg}$ \citep{nomoto_84,nomoto_87,jones_13,jones_14,jones_19,leung_19,leung_20} (although collapse is not certain \citep{kirsebom_19}),  different from  iron core-collapse supernovae that undergo further hydrostatic burning stages and collapse only after iron core formation. For electron-capture supernovae, a uniform core structure and mass are set by the threshold
core mass to start electron capture and ignite O-Ne-deflagration
\citep{leung_20}, and therefore a unique neutron star mass is expected.  Accretion-induced collapse in binary stars with an O-Ne-Mg white dwarf is an alternative avenue to EC supernovae.  Similarly low neutron star masses were obtained in simulations of the collapse of low-mass iron core progenitors \citep{mueller_12b,melson_15a,mueller_19a}.
Recent simulations have in fact pushed the lower
limit for predicted baryonic neutron star masses
down slightly to  $1.347\msun$ \citep{burrows_24}.
Even lower values were reported in 2D simulations \citep{suwa_18}, but their supernova simulations
were performed with simplified neutrino transport for stellar evolution models based on artificial bare C/O cores with unrealistically low final C/O core masses $<1.45\msun$ for such objects. This introduces an artificial bias in the remnant masses. Fixing the C/O core mass eliminates
essential processes for core growth or core shrinkage by dredge-up, dredge-out \citep{nomoto_87,doherty_15} and mass transfer that determine the final fate and the ultimate core sizes at the boundary between SAGB stars and massive stars. Even for ultra-stripped stars that lose much of their He shell by binary interaction, further C/O core growth after the ignition of C burning is important, and brings the final C/O core mass to $\gtrsim 1.45 \msun$ in stars that evolve to iron core collapse.

This tension between observations and theory leaves open several interpretations. Are stellar evolution models not capturing the core structure of the least massive supernova progenitors correctly? Do supernova explosions develop faster than predicted in current models to allow a smaller mass cut? Or are the observed low-mass objects indeed white dwarfs instead of neutron stars \citep{tauris_19}?

One key priority for supernova simulations in answering these questions is to better scan the mass range of progenitors that produce the lightest neutron stars. Even though a number of supernova simulations of electron-capture supernovae and low-mass iron core collapse supernovae have been carried out \citep{kitaura_06,janka_08,mueller_12b,mueller_13,melson_15a,mueller_16b,radice_17,gessner_18,mueller_18,stockinger_20,zha_22},
relatively few progenitor models are available to date due to technical difficulties that plague pre-collapse stellar evolution in this region of parameter space. The final stages of the progenitors in this regime are
characterized by highly degenerate conditions in and around the core and complex burning behavior that can involve off-center ignition, convectively bounded flames, and sometimes powerful burning flashes with pre-supernova mass ejection \citep{woosley_15}. Crucially, the core structure does not depend monotonically on initial progenitor mass \citep{mueller_16a}. This means that low-mass iron core collapse supernovae could form lighter neutron stars than electron capture supernovae despite their higher ZAMS mass. Similarly a star with slightly higher ZAMS mass can may sometimes make a smaller iron core and neutron star.
A fine grid of stellar evolution models is required to scan the variations in core size just above the iron-core formation threshold.

In this paper, we improve the theoretical lower limit for the neutron star mass compatible with current stellar evolution and supernova explosion theory. Starting from a finely spaced grid of low-mass iron-core progenitor models, we select five suitable candidates for particularly low neutron star masses. We then simulate the collapse and, if applicable, the explosion of these progenitor stars in three dimensions (3D) and predict the associated neutron star birth properties.

\begin{figure}
    \centering
    \includegraphics[width=\linewidth]{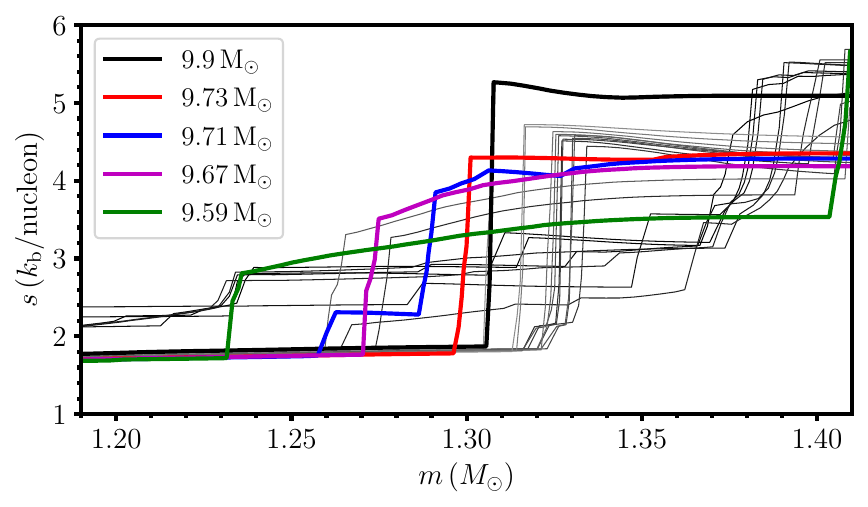}
    \caption{Entropy profiles of all 25 low-mass iron core progenitor models near the potential mass cut. Five models that have been identified as suitable candidates for three-dimensional supernova simulations are shown as thick colored lines. The remainder of the set are shown as thin gray lines, with lighter shades indicating higher ZAMS mass.}
    \label{fig:progs}
\end{figure}

\textit{Progenitor Models.}---We consider 25 single-star solar-metallicity progenitor models covering the ZAMS mass range from
$9.45\msun$ to $9.95\msun$
and C/O core masses from $1.481\msun$ to $1.585\msun$.  The models have been calculated with the stellar evolution code \textsc{Kepler} \citep{weaver_78,heger_10}, including fixes to the pair neutrino rates as applied previously in \citep{mueller_16a,SWH18}.
The resolution in ZAMS mass is as small as $0.01\msun$ where possible, but due to the complications of highly degenerate core evolution, in particular at the low-mass end, only 25 models were followed until collapse.  In the early evolution stages, a dynamic adaptive nuclear reaction network is used that follows all necessary stable and unstable isotopes up to polonium (typically, some 1,800 species for the pre-supernova model). Silicon burning and evolution of the iron core use a $\mathord\sim125$ isotope network for intermediate and full nuclear statistical equilibrium for efficient advection of conserved quantities while reasonably capturing deleptonization and neutrino emission as well as energy in the nuclear excited states \citep{weaver_78,woosley_15}.

Due to the computer time requirements of 3D core-collapse supernova simulations with neutrino transport, a subset of five progenitor models was selected for follow-up. Suitable candidates for a low neutron star mass were identified based on the entropy profiles of the progenitors (Figure~\ref{fig:progs}). The onset of the explosion is typically associated with entropy and density jumps at shell interface, in particular the pronounced entropy jump at the
base of the oxygen-burning shell in many models 
\citep{janka_12b,mueller_14,ertl_15,vartanyan_18,wang_22}. Figure~\ref{fig:progs} shows the biggest entropy jump for the $9.90\msun$ model at 
just above $1.3\msun$. Sizable entropy jumps at smaller mass coordinate $m$ occur, but are less prominent, and may not be sufficient to trigger an explosion.  We select the $9.9\msun$, 
$9.73\msun$, $9.71\msun$, $9.67\msun$,
and $9.59\msun$ progenitors as five candidates with progressively smaller iron core masses of $1.31\msun$,
$1.30\msun$, $1.29\msun$,
$1.27\msun$, and $1.23\msun$, but progressively lower likelihood of an explosion triggered by the entropy jump.

Of particular interest is the final evolution that leads to the small core sizes.  Post helium burning, all models undergo 4 episodes of carbon core and shell burning before neon/oxygen burning is ignited off-center at $\mathord\sim0.19\msun\ldots0.32\msun$ for the most to the least massive models and burns all the way downward to the center.  The $9.9\msun$ model first ignites core silicon burning slightly off-center, the other models ignite silicon burning directly in the center.  All models encounter an oxygen shell burning phase before igniting silicon shell burning.  In the $9.90\msun$ and $9.73\msun$ models, it is quenched well before silicon shell ignition, and sufficient oxygen remains to re-ignite and drive a powerful convective oxygen burning shell minutes ($9.90\msun$) to seconds ($9.73\msun$ and below) before core collapse just at the outer edge of the iron core, giving rise to the huge jumps in entropy seen in Figure~\ref{fig:progs}.  In the lower-mass models, oxygen is increasingly depleted in the pre-silicon burning phase of oxygen shell burning such that the that pre-collapse oxygen shell results in smaller entropy jumps.

\textit{Numerical Methods.}---We perform three-dimensional supernova simulations with the
relativistic neutrino hydrodynamics code \textsc{CoCoNuT-FMT} code \citep{mueller_15a}.
\textsc{CoCoNuT-FMT} solves the relativistic
equations of hydrodynamics in spherical polar coordinates using the piecewise
parabolic method \citep{colella_84} for reconstruction
and a hybrid HLLC/HLLE Riemann solver \citep{mignone_05_a,harten_83}. 
A mesh coarsening scheme is used to avoid time step constraints near the axis of the spherical polar grid \citep{mueller_19a}, and the innermost region of the grid is treated in spherical symmetry.
Neutrinos are treated with the fast multi-group transport (\textsc{FMT}) scheme of \citep{mueller_15a}, and the equations for the space-time metric are solved in the extended conformal flatness approximation \citep{cordero_09}.
We use a grid resolution of $550\times128\times 256$
zones in radius, latitude, and longitude (corresponding
to $1.4^\circ$ in angle)  and $21$ exponentially spaced energy group from $4 \, \mathrm{MeV}$
to $240 \, \mathrm{MeV}$. We use the SFHo equation of state of
\citep{steiner_13} in the high-density regime. At low densities, the equation of state includes nucleons and nuclei (treated as a perfect gas), leptons and radiation, and nuclear burning is treated with a ``flashing'' scheme \citep{rampp_02}.

\begin{figure}
    \centering
    \includegraphics[width=\linewidth]{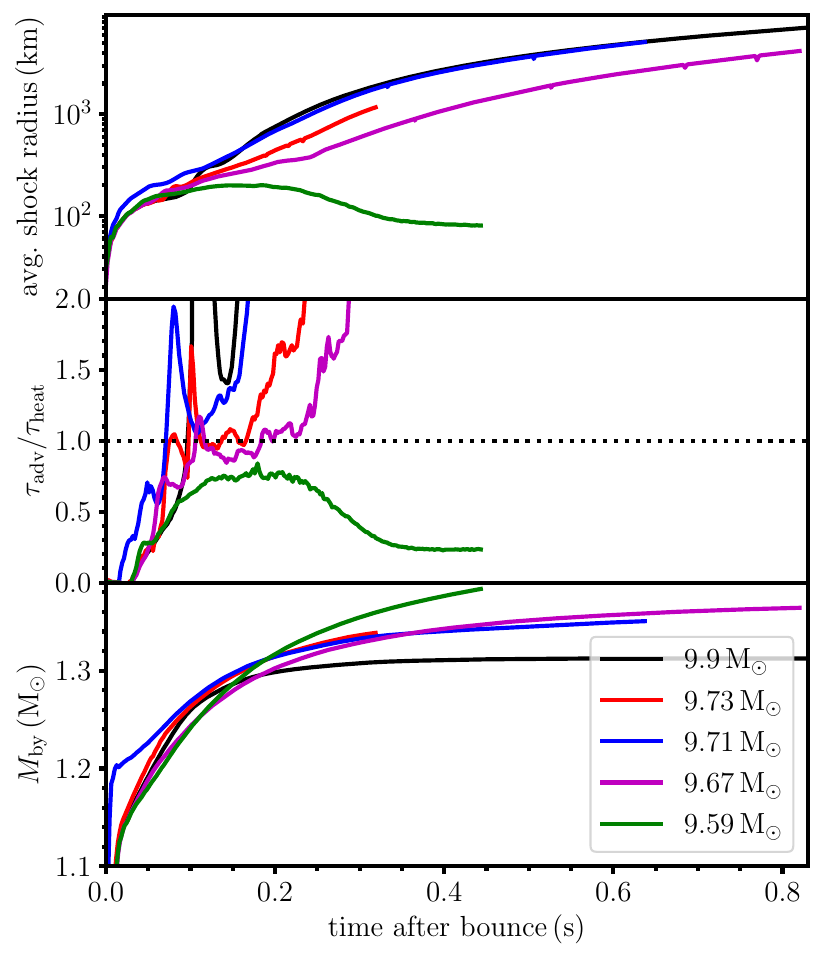}
    \caption{Average shock radius (top),
    criticality ratio $\tau_\mathrm{adv}/\tau_\mathrm{heat}$ between
    the advection and heating time scale,
    and baryonic neutron star mass (bottom) as a function of time for the five core-collapse supernova simulations.}
    \label{fig:expl}
\end{figure}

\textit{Results.}---Shock trajectories for the five models are shown in Figure~\ref{fig:expl}. The $9.9\msun$ 
and $9.71\msun$ models are the earliest to explode. The  $9.59\msun$ model fails to explode early when the shell interface at $1.23\msun$ reaches the shock, though it may still explode later, perhaps triggered by the more pronounced entropy jump at $1.4\msun$.
The critical threshold for shock revival by neutrinos is often expressed via the ratio between the advection time scale
$\tau_\mathrm{adv}$
and the heating time scale $\tau_\mathrm{heat}$ \citep{buras_06a,mueller_12a}.
This ratio quantifies whether neutrino heating imparts sufficient energy on accreted material before it is advected from the shock to the bottom of the heating region.
The critical ratio $\tau_\mathrm{adv}/\tau_\mathrm{heat}$ shows
the same hierarchy, with the $9.71 \msun$ model reaching $\tau_\mathrm{adv}/\tau_\mathrm{heat}=1$ the earliest. Even though 
the $9.67 \msun$, $9.71 \msun$,
and $9.73 \msun$ models initially have slightly higher values for a brief period,
the $9.9 \msun$ simulation eventually shows the fastest
shock propagation, followed very closely
by the  $9.71 \msun$ case.  Evidently, the more rapid drop of the accretion rate associated with the bigger entropy jump in the $9.9 \msun$ model is more critical for the rapid development of an explosion than the smaller mass coordinate of the jump in the $9.67 \msun$ and $9.73 \msun$ models.

By the end of the simulation, the $9.9\msun$ model has reached an explosion energy of $1.5\times 10^{50}\,\mathrm{erg}$, which is significantly higher than for the $9.67\msun$ model with $4.7\times 10^{49}\,\mathrm{erg}$, and somewhat higher than for the $9.71\msun$ model
at $10^{50}\,\mathrm{erg}$. The 
 $9.73\msun$ model has the lowest energy by the end of the simulation, but has been terminated earlier than the other simulations and lies between the $9.67\msun$ and the  $9.71\msun$ case by the end of the simulation.

The key outcome for the purpose of this paper is the neutron star mass. Figure~\ref{fig:expl} (bottom panel) shows the evolution of the baryonic neutron star mass, $M_\mathrm{by}$, which is obtained by integrating the entire mass on the grid at densities above $10^{11}\,\mathrm{g}\,\mathrm{cm}^{-3}$.
Reflecting the more precipitous explosion in the $9.9\msun$ model, the neutron star mass quickly saturates at $1.313\msun$ in this case, whereas $M_\mathrm{by}$ still continues to increase in the other three exploding models. By the end of the simulation, the neutron star is already losing mass at a very small rate in the $9.9\msun$ model, so the value of
$M_\mathrm{by}=1.313\msun$ is a rather firm upper limit for the final neutron star mass, barring the possibility of later fallback, which would be unlikely for a stripped progenitor in system J0453+1559.

\begin{figure}
    \centering
    \includegraphics[width=\linewidth]{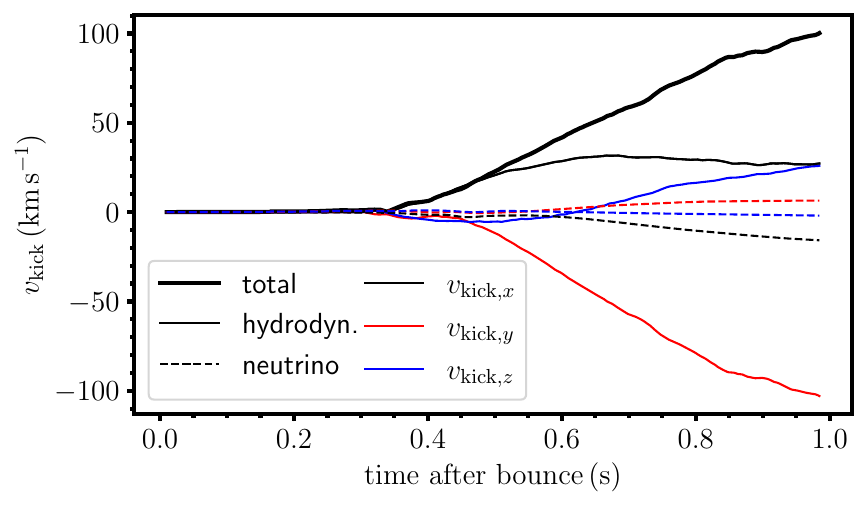}
    \caption{Total kick velocity (thick solid line)
    and the $x$- (thin black lines), $y$- (red), and $z$-components (blue) of the kick due
    to the gravitational tug-boat mechanism (hydrodynamic kick, thin solid lines) and
    due to anisotropic neutrino emission (dashed lines) for the $9.9\msun$ model.}
    \label{fig:kick}
\end{figure}

Using the fit formula for the cold neutron star binding energy from \citep{lattimer_01}, $M_\mathrm{by}$ can be translated into the final gravitational mass $M_\mathrm{grav}$,
\begin{equation}
    M_\mathrm{by}=M_\mathrm{grav}-
    0.084 \msun \left(M_\mathrm{grav}/\mathrm{M}_\odot\right)^2,
\end{equation}
which results in $M_\mathrm{grav}=1.192 \msun$. This is still in conflict with the mass measurement of $M_\mathrm{grav}=1.174\msun$ for the lighter companion in  J0453+1559. If we assume
$2\,\sigma$ error bars instead of $1\,\sigma$ error bars, the results of \citep{martinez_15} may be compatible with a mass as high as $1.182\msun$, but even then a tension with observations remains. It is conceivable that stripped stars may have even slightly more favorable structure for low-mass neutron stars as their expected higher carbon mass fraction at the end of core helium burning may lead to more neutronisation in the supernova progenitor \citep{woosley_19}. Something similar may occur for more metal-rich progenitors. (see \textit{Conclusions}).

Compatibility with the system parameters of J0453+1559, specifically the eccentricity and orbital period, also requires a suitable kick and amount of mass loss in the explosion. 
A small kick, as typically expected for electron-capture supernovae, is required \citep{martinez_15}. For their alternative scenario involving
the formation of a massive white dwarf and the
ejection of several $0.1\msun$ in a \emph{thermonuclear} electron-capture supernova,
Tauris \& Janka \citep{tauris_19} estimate a required kick of $\mathord{\sim} 70\,\mathrm{km}\,\mathrm{s}^{-1}$. With a single-star progenitor we cannot estimate the realistic amount of mass ejection in a binary scenario where the progenitor would have undergone envelope stripping earlier in its evolution. It is plausible, however, that the resulting envelope mass in a binary scenario and the required kick will be of a similar scale in case of formation of the light companion by a low-mass iron core progenitor \citep{tauris_15,mueller_19a}.

We therefore evaluate the kick for the $9.9\msun$ model to demonstrate that the model is roughly consistent with these requirements with a kick of order $\mathord{\sim}100\,\mathrm{km}\,\mathrm{s}^{-1}$. Since the kick velocity is subject to stochastic model variations by at least a factor of two in both directions, only the rough scale of the kick is relevant. 
Since the simulation uses an immobile spherical core of about $7.5\,\mathrm{km}$,
we compute the neutron star momentum indirectly from the vectorial momentum of the \emph{ejecta} $\mathbf{p}_\mathrm{ej}$, which arises due
to the gravitational tug-boat mechanism \citep{scheck_06}, and the emitted neutrinos by invoking momentum conservation \citep{scheck_06,wongwathanarat_13}.
In terms of the relativistic momentum density $\mathbf{S}$ and the neutrino energy flux $\mathbf{F}$ (measured at a radius of $400\,\mathrm{km}$), we compute the kick velocity $\mathbf{v}_\mathrm{kick}$ at time $t$ as
\begin{equation}
    M_\mathrm{grav} \mathbf{v}_\mathrm{kick}(t)
    =-\int\limits_\mathrm{ejecta} \mathbf{S}\,\ud V
    -\int\limits_0^t\!\!\oint \frac{\mathbf{F}(r=400\,\mathrm{km})}{c}\,\ud A,
\end{equation}
where the relativistic volume and surface elements are to be used. This procedure for post-processing simulations circumvents the issue of numerical non-conservation of momentum and yields similar results to simulations using  an accelerating reference frame \citep{scheck_06}  or without a fixed core \citep{nordhaus_12}.
By the end of the simulation, the kick velocity $|\mathbf{v}_\mathrm{kick}|$ has reached a value of
$100\,\mathrm{km}\,\mathrm{s}^{-1}$ and is still growing, albeit at a decelerating pace.
The kick from asymmetric neutrino emission is subdominant, but slightly reduces the total kick as it points in the direction opposite to the hydrodynamic kick. The $9.71 \msun$ model shows a similar growth of the kick velocity, which demonstrates that kicks of this order are not predicted to be unusual in this progenitor mass range.

\textit{Conclusions.}---
Our simulations set a new record for the lowest
neutron star mass obtained in 3D supernova simulations with multi-group neutrino transport.
The lowest gravitational neutron star mass compatible with current stellar evolution models is 
now at least as low as $1.192\msun$. Despite
the rapid explosion of the
$9.9\msun$  progenitor and the
small amount of accretion after shock revival,
we are able to obtain a kick of $100\,\mathrm{km}\,\mathrm{s}^{-1}$, which is mostly due to the gravitational tug-boat mechanism \citep{scheck_06} with a small additional contribution from anisotropic neutrino emission. There is still a small tension with the $2\,\sigma$ error bars for
the $1.174\msun$ object in  J0453+1559, whose neutron star nature has recently been questioned
\citep{tauris_19}. The tension is so small, however, so that minor systematic or stochastic variations in the progenitor evolution and supernova dynamics may well
extended the range of neutron star masses down by 
another $\mathord{\sim} 0.01 \msun$ and resolve this discrepancy. Our progenitor set may have not yet sampled the
very extremes of the stochastic variations in progenitor structure.
Furthermore, convective seed perturbations in the progenitor could result in a slightly earlier onset of the explosion
\citep{mueller_15b,couch_15,mueller_17} and reduce the neutron star mass further. The intricacies of  degenerate nuclear burning may well introduce uncertainties in the pre-collapse core structure that also result in slightly smaller neutron star masses. Even minor details in the pre-collapse models, such as the usual non-rigorous treatment of the small effect of relativistic gravity in current stellar evolution code, or
inward growth of the oxygen shells by entrainment \citep{rizzuti_22}, could affect the pre-collapse evolution on a level of accuracy that matters for the comparison to J0453+1559 and other future neutron star candidates with precise mass determination in this mass range. The nuclear equation of state could also have a minor impact.

The neutron star interpretation has the advantage that the requisite kick emerges naturally from a 3D simulation, whereas the white dwarf interpretation \citep{tauris_19} relies on an optimistic estimate for the ejecta momentum anisotropy of $\alpha\sim 0.05$. The ashes in thermonuclear electron capture supernovae only have mild global asymmetries \emph{and} the density contrast between fuel and ashes will be low because the turbulent flame is subsonic with Mach number $\sim 0.05$ \citep{jones_16}, suggesting a much smaller anisotropy factor.

Regardless of whether the $1.174\msun$ object in  J0453+1559 can eventually be explained as a neutron star, our new simulations hold several important insights. 
First, our new mass limit also relieves tensions between
3D models and several other neutron stars with low masses of $1.21\texttt{-}1.22\msun$ \citep{you_24}.
Second, contrary to long-standing expectations, the lightest neutron star masses appear not to be made by electron-capture supernovae, but by iron-core collapse supernovae. Furthermore, models predict considerable non-monotonicity in the explosion and remnant properties for iron-core collapse supernovae near the minimum progenitor mass. The lightest neutron stars may originate from
stars several $0.1\msun$ above the progenitor mass threshold (in terms of ZAMS mass) for core-collapse supernovae. Furthermore, there is no strict correlation between neutron star masses and kicks. We predict that the lightest neutron stars can still have substantial (albeit below-average) kicks.
Combining increasingly detailed and precise
predictions from 3D explosion models and
more precise neutron star mass and kick determinations at the low-mass end of the distribution holds considerable promise for testing
and validating our understanding of supernova physics and stellar evolution in this challenging progenitor mass regime.

\textit{Data availability}---
The data from our simulations will be made available upon reasonable requests made to the authors. 

\textit{Acknowledgments}--- The authors acknowledge support by the Australian Research Council through grants FT160100035 (BM) DP240101786 (BM, AH) DE210101050 (JP) and DP240103174 (AH).
The authors acknowledge support
by the Australian Research Council (ARC) Centre of Excellence (CoE) for Gravitational Wave Discovery (OzGrav) project numbers CE170100004 and CE230100016. 
This project was carried out using computer time allocations from Astronomy Australia Limited's ASTAC scheme, the National Computational Merit Allocation Scheme (NCMAS), and
from an Australasian Leadership Computing Grant.
Some of this work was performed on the Gadi supercomputer with the assistance of resources and services from the National Computational Infrastructure (NCI), which is supported by the Australian Government, and through support by an Australasian Leadership Computing Grant.

\bibliographystyle{h-physrev}
\bibliography{paper}

\label{lastpage}

\end{document}